\RequirePackage{lineno}
\documentclass[aps,prb,twocolumn,superscriptaddress]{revtex4}
\usepackage{float}
\usepackage{amsmath}
\usepackage{amssymb}
\usepackage[utf8x]{inputenc}
\usepackage[T1]{fontenc}
\usepackage{mathptmx}
\usepackage{braket}
\usepackage{graphicx}
\usepackage{epstopdf}
\usepackage[amssymb]{SIunits}
\setcitestyle{super}  %% For superscript citations
\usepackage[colorlinks=true, linkcolor=blue, urlcolor=blue, citecolor=blue, plainpages=false, pdfpagelabels, bookmarksnumbered]{hyperref}
\usepackage{todonotes}
\usepackage{etoolbox}
\patchcmd{\section}
  {\centering}
  {\raggedright}
  {}
  {}
\patchcmd{\subsection}
  {\centering}
  {\raggedright}
  {}
  {}

\def\XXint#1#2#3{{\setbox0=\hbox{$#1{#2#3}{\int}$}
     \vcenter{\hbox{$#2#3$}}\kern-.5\wd0}}

\newenvironment{methods}{%
    \setlength{\parindent}{0in}%
    \section*{Methods}%
    \setlength{\parskip}{12pt}%
    }{}

\newenvironment{addendum}{%
    \setlength{\parindent}{0in}%
    \small%
    \begin{list}{Acknowledgements}{%
        \setlength{\leftmargin}{0in}%
        \setlength{\listparindent}{0in}%
        \setlength{\labelsep}{0em}%
        \setlength{\labelwidth}{0in}%
        \setlength{\itemsep}{12pt}%
        }
    }
    {\end{list}\normalsize}

\newcommand{\mybibliography}[2]{\IfFileExists{#1}{\bibliography{#1}}{\bibliography{#2}}}
\begin{document}

\title{Controlled free-induction decay in the extreme ultraviolet}

\makeatletter

\author{S.~Bengtsson}
\thanks{These authors contributed equally.}
\affiliation{Department of Physics, Lund University, P.O. Box 118, SE-221 00 Lund, Sweden.}
\author{E.~W.~Larsen}
\thanks{These authors contributed equally.}
\affiliation{Department of Physics, Lund University, P.O. Box 118, SE-221 00 Lund, Sweden.}
\author{D.~Kroon}
\affiliation{Department of Physics, Lund University, P.O. Box 118, SE-221 00 Lund, Sweden.}
\author{S.~Camp}
\affiliation{Louisiana State University, Baton Rouge, 70803-4001, Louisiana, United States of America.}
\author{M.~Miranda}
\affiliation{Department of Physics, Lund University, P.O. Box 118, SE-221 00 Lund, Sweden.}
\author{C.~L.~Arnold}
\affiliation{Department of Physics, Lund University, P.O. Box 118, SE-221 00 Lund, Sweden.}
\author{A.~L'Huillier}
\affiliation{Department of Physics, Lund University, P.O. Box 118, SE-221 00 Lund, Sweden.}
\author{K.~J.~Schafer}
\affiliation{Louisiana State University, Baton Rouge, 70803-4001, Louisiana, United States of America.}
\author{M.~B.~Gaarde}
\affiliation{Louisiana State University, Baton Rouge, 70803-4001, Louisiana, United States of America.}
\author{L.~Rippe}
\affiliation{Department of Physics, Lund University, P.O. Box 118, SE-221 00 Lund, Sweden.}
\author{J.~Mauritsson.}
\affiliation{Department of Physics, Lund University, P.O. Box 118, SE-221 00 Lund, Sweden.}
\makeatother

\begin{abstract}

Coherent sources of attosecond extreme ultraviolet (XUV) radiation present many challenges if their full potential is to be realized. While many applications benefit from the broadband nature of these sources, it is also desirable to produce narrow band XUV pulses, or to study autoionizing resonances in a manner that is free of the broad ionization background that accompanies above-threshold XUV excitation. Here we demonstrate a method for controlling the coherent XUV free induction decay that results from using attosecond pulses to excite a gas, yielding a fully functional modulator for XUV wavelengths. We use an infrared (IR) control pulse to manipulate both the spatial and spectral phase of the XUV emission, sending the light in a direction of our choosing at a time of our choosing. This allows us to tailor the light using opto-optical modulation, similar to devices available in the IR and visible wavelength regions.

\end{abstract}
\maketitle
%\section{Format requirements}

%\section{Instructions}
%Title page: a title of 15 words or fewer, without punctuation; author
%affiliations and contact information (the corresponding author should
%be identified with an asterisk).
%
%For Letters: a referenced introductory paragraph of approximately
%150 words; main text not to exceed 2,000 words; no more than 3–5
%display items (figures, tables); references are limited to 30.
%
%\textbf{For Articles}: an abstract of approximately 150 words, unreferenced;
%main text (excluding abstract, Methods, references and figure legends) of 2,000–3,000 words; no more than 4–6 display items
%(figures, tables); references are limited to 50. An introduction of up to 500 words of referenced text expands on the background of the work (some overlap with the summary is acceptable), followed by a concise, focused account of the findings, ending with one or two short paragraphs of discussion.
%
%\textbf{Methods}: authors may supply a section of up to 800 words
%describing key methods. Additional information may be placed in
%Supplementary Information.

The ability to control light is central to any optical application. Though the methods used  depend on the spectral regime,  they have in common that both the spectral and spatial phase are controlled. This phase control can be exerted for instance by the acousto-optic effect, where a sound wave is sent through a crystal~\cite{DebyePNAS1932}; the electro-optic effect, using a slow electric field to change the light-matter response; or by the use of a spatial light modulator~\cite{weinerRSI2000}. These techniques are generally used to modulate light with frequencies that can be sent through a crystal or a fiber~\cite{KaoRMP2010}, and they are therefore limited to light with visible or longer wavelengths, due to absorption. The need for methods to control the spatial and spectral properties of light with shorter wavelengths has been manifested by the recent development of both attosecond extreme ultraviolet (XUV) sources~\cite{PaulScience2001,HentschelNature2001,KrauszRMP2009} from high harmonic generation (HHG)~\cite{McPhersonJOSAB1987,FerrayJPB1988,SchaferPRL1993,CorkumPRL1993,LewensteinPRA1994} and soft x-ray pulses from free electron lasers (xFELs)~\cite{AckermannNP2007,EmmaNP2010}. These sources of ultra-broadband, coherent radiation are making possible time-resolved measurements of electron dynamics in atoms, molecules, and solids~\cite{DrescherNature2002,KlingScience2006,StockmanNPho2007,CavalieriNature2007,SansoneNat2010,FangPRL2010,CryanPRL2010,GlowniaOE2010,YoungNat2010,GoulielmakisNat2010,DoumyPRL2011,klunderPRL2011}.

We propose to use the phenomena of controlled free induction decay (FID) to extend the possibilities of phase and amplitude control to the XUV. FID is the emission that follows any coherent excitation of an ensemble of atoms in the absence of an external electromagnetic driving field, and was first proposed~\cite{BlochPR1946} and seen~\cite{HahnPR1950} in nuclear magnetic resonances (NMR)  in the late 1940s, and later observed in the visible regime~\cite{brewerPRA1972,HopfPRA1973}. The emission is due to the excitation of resonant states and it has the same spatial properties as the excitation pulse, but with a phase-shift. The interference between the two fields at the resonant frequencies yields the  absorption spectrum observed in optical spectroscopy. If, for example, the individual emitters are $\pi$ out of phase with the excitation pulse then the normal Lorentzian lineshape is observed. For any other phase there will be an asymmetric Fano profile~\cite{Fano1961PR}, as for example when an autoionizing resonance is excited. Control over the phase of individual  XUV emitters and the consequent change in the absorption lineshape by  XUV FID (xFID) has been 
demonstrated  in  attosecond transient absorption experiments in gas-phase atoms and molecules~\cite{WangPRL2010, OttNature2014,BeckNJP2014,LiaoPRL2015,wuPRA2013}, and in metals~\cite{Vura-weisPCL2013}. Beaulieu~\textit{et al.} also showed the long decay time of this emission~\cite{BeaulieuPRL2016}.

In this article we extend the temporal control of xFID to the spatial domain thus demonstrating an opto-optical modulator for XUV light. We use an infrared (IR) control pulse to manipulate both the spatial and spectral properties of resonant emission in the XUV, sending the emission in a direction of our choosing at a time of our choosing.  By adding control over the spatial phase of the xFID emission we can both create narrow-band sources of XUV radiation tuned to specific frequencies and we can study resonant emission in a background-free measurement scheme. Our opto-optical modulator also provides a pathway toward synchronizing x-ray from xFELs and IR pulses at the sub-femtosecond level -- an order of magnitude better than what is presently possible. 
% In addition to its practical aspects, our work also demonstrates the benefits of thinking about absorption and emission in the time domain as a field-driven process, rather than in the frequency domain using a photon-driven picture.
% Alternate version
Our work also demonstrates the benefits of thinking about absorption and emission as field-driven processes in the time domain, which complements the more common picture of photon-driven processes in the frequency domain.

\begin{figure*}
\includegraphics[width=\textwidth]{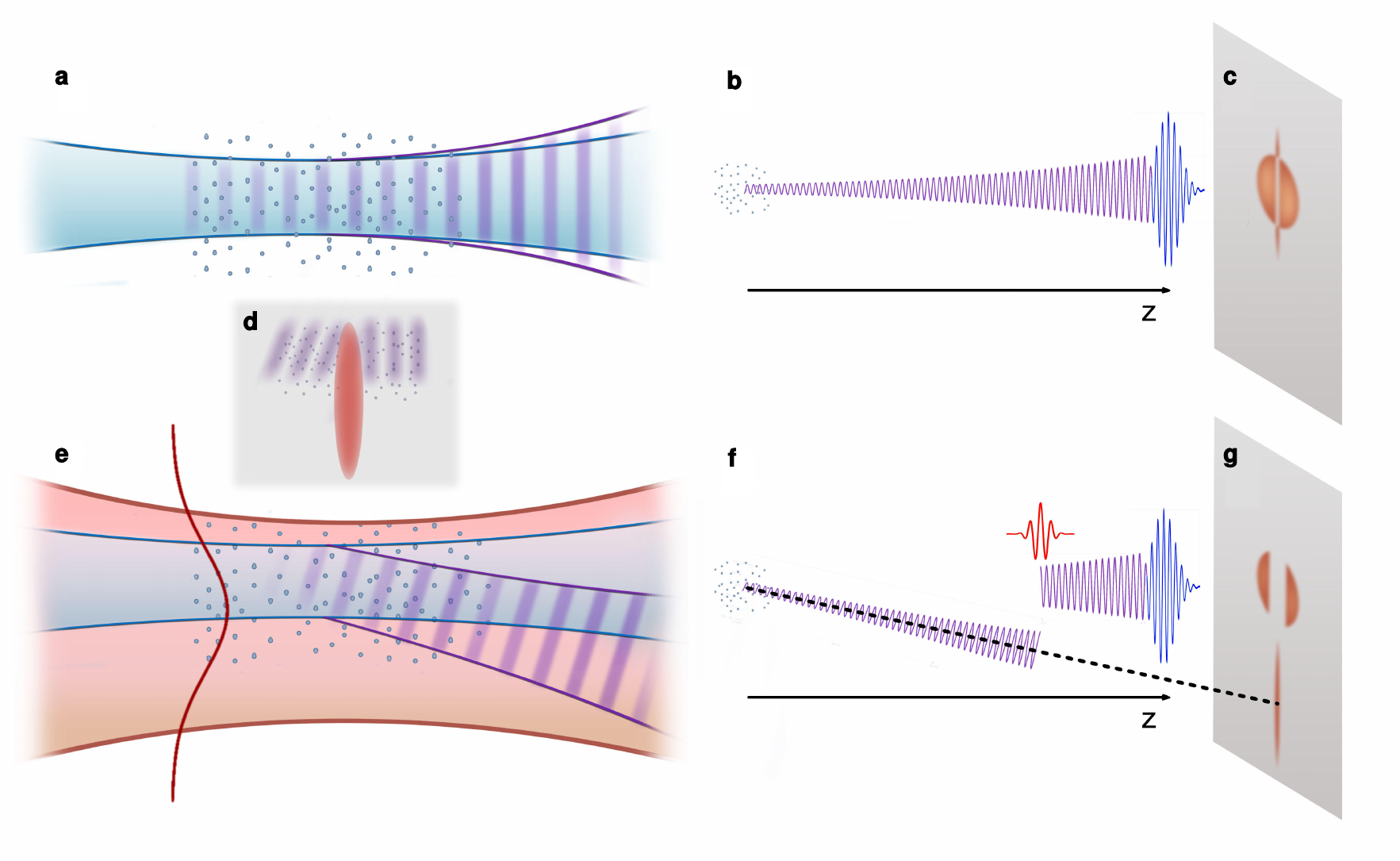}
\caption{\textbf{Schematic illustration of xFID radiation control.} 
\textbf{a} An ensemble of atoms excited by an ultrafast XUV pulse emit xFID radiation (purple) after the XUV pulse (blue) has passed. The atoms oscillate in phase orthogonal to the propagation direction of the excitation pulse (illustrated by the vertical purple lines).  Phase matching creates a well-defined xFID beam. \textbf{b} The temporal structure of the xFID emission decays over a long time. \textbf{c} The phase relation between the excitation pulse and the xFID leads to destructive interference at the detector in the far-field, normally called absorption. \textbf{d} An IR pulse (red) that co-propagates with the XUV pulse through the  medium creates a spatial-dependent phase shift of the dipoles via the AC-Stark effect. This phase shift depends on the integrated IR intensity for each atom and results in a rotation of the wave fronts after the IR pulse. \textbf{e} This rotation redirects the xFID emission after the IR pulse. \textbf{f, g} In the far field this yields an off-axis emission component and an altered on-axis absorption as well.}
\label{fig:overviewscheme}%
\end{figure*}

\begin{figure}[tbh]
\includegraphics[width = \columnwidth]{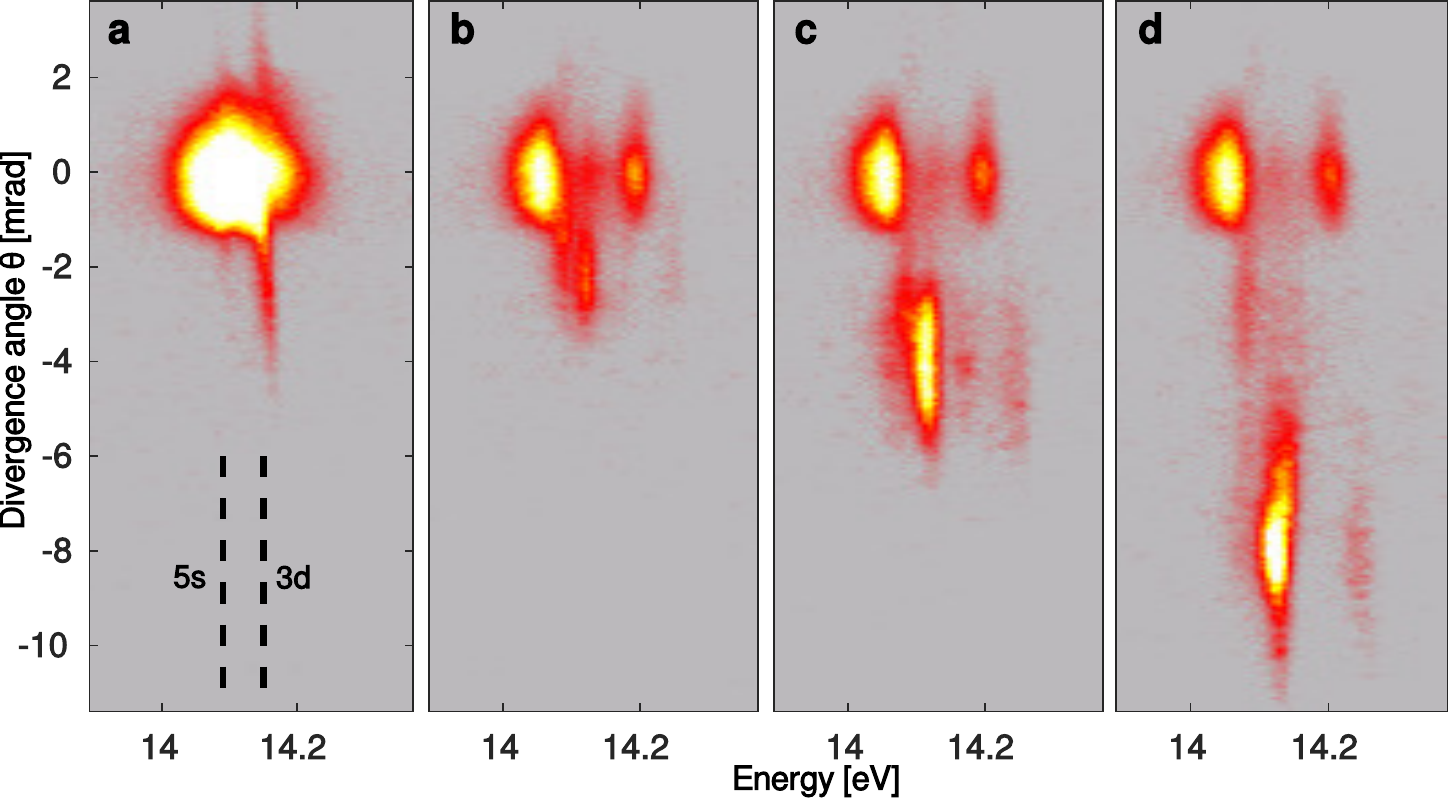} %2015-04-30 probe(1 34 32 31)
\caption{\textbf{Experimental spatial--spectral profile of the VUV pulse consisting of the 9th harmonic of 780\,nm after transmission through argon.} \textbf{a} Without an IR control pulse. \textbf{b} An off-centered IR pulse follows the harmonics through argon and redirected, off-axis emission is seen from the states resonant with harmonic 9. \textbf{c}-\textbf{d} By increasing the IR intensity the induced wavefront rotation increases and therefore also the redirection angle.}
\label{fig:intensity}
\end{figure}

To induce the xFID we use a tunable XUV pulse produced by HHG~\cite{McPhersonJOSAB1987,FerrayJPB1988,SchaferPRL1993,CorkumPRL1993,LewensteinPRA1994} to create a coherent superposition of ground and excited states in gas-phase argon. After being coherently excited, the different atoms along the propagation direction emit with a fixed phase relation. This leads to phase-matching of the dipole emission in the forward direction, resulting in an xFID signal~[Fig. \ref{fig:overviewscheme}~\textbf{a-c}]. Though the xFID is emitted in the same direction as the excitation pulse, the temporal properties of the emission depend on the coherence time of the excited state. This results in an exponentially decaying xFID emission following the excitation pulse, which allows interaction between the excited atoms with an additional light pulse~[Fig.~\ref{fig:overviewscheme}~\textbf{d-g}], enabling phase control without perturbing the excitation processes. When a moderately strong IR control pulse interacts with the excited atoms in the ensemble the energy levels will be AC-Stark shifted. If the laser pulse is not too strong (\textit{i.e.} not ionizing the medium), the energy levels will follow the field adiabatically and return to their initial energy after the pulse, with the only difference being that the superposition has accumulated an additional phase shift $\Delta \varphi$. This phase shift depends on the intensity of the IR pulse:
\begin{equation}
\Delta \varphi(x,y,z) = \int_{\textrm{T}_{\textrm{IR}}} {\frac{\Delta E(x,y,z,t)}{\hbar} \textrm{dt} },
\label{eq:phaseshift}
\end{equation}
where $\Delta E$ is the intensity-dependent energy shift of the levels and $\textrm{T}_{\textrm{IR}}$ the duration of the IR pulse. If all the atoms are exposed to the same time-dependent IR intensity, the induced phase shift will lead to interference between the part of the xFID that comes before the IR pulse and the part that comes after, thereby modifying the spectral and temporal structure of the emission in the forward direction. By parallel shifting the IR beam, different atoms are exposed to different IR intensities, imposing a spatial phase variation across the gas. This allows us to tailor the wavefront of the xFID, redirecting all the emission that comes after the control pulse [see schematic illustration in Fig.~\ref{fig:overviewscheme}~\textbf{d}-\textbf{g} and experimental demonstration in Fig.~\ref{fig:intensity}].

Our experimental technique (see Methods and Fig.~\ref{fig:experimentalapparatus}) utilizes an attosecond transient-absorption scheme~\cite{GoulielmakisNat2010,WangPRL2010,OttScience2013} in a non-concentric geometry together with a high resolution, flat-field imaging spectrometer. In the first set of measurements that we will discuss, the laser wavelength is 780\,nm so that the 9th harmonic is centered at 14\,eV and the broadband VUV light spectrally covers both the 3s$^2$3p$^6\rightarrow$3s$^2$3p$^5$5s (14.09\,eV) and 3s$^2$3p$^6\rightarrow$3s$^2$3p$^5$3d (14.15\,eV) transitions. In Fig.\,\ref{fig:intensity} experimental results for different IR intensities are presented. When no  IR control pulse is used [Fig.\,\ref{fig:intensity}\,\textbf{a}] the xFID signal consists of narrow spectral lines that are slightly more divergent than the original pulse. 
The excited states have long coherence times, which leads to narrow spectral features, while the dephasing that takes place during this time results in a phase front distortion and increased divergence. 
In Fig.\,\ref{fig:intensity}\,\textbf{b}-\textbf{d} we use an IR control pulse that comes 200\,fs after  the excitation pulse.  As the control pulse interacts with the medium it induces a spatial phase variation across the ensemble of excited atoms. This terminates the on-axis component of the xFID field, and the subsequent xFID signal is redirected, creating the observed off-axis signal. With increasing intensity the spatial phase variation increases and the redirection angle increases as a result.

The on- and off-axis components of the xFID also display different spectral characteristics. The total on-axis signal results from coherent superposition of the incoming field and the generated xFID field in the forward direction. Since the xFID emission is $\pi$ phase-shifted with respect to the excitation pulse, the two fields destructively interfere. This results in a spectral hole at the transition energies, normally called absorption. Since the xFID emission in the forward direction is terminated by the IR control pulse, thus effectively reducing the lifetime in this direction, the spectral width of the on-axis absorption will increase with respect to the no IR case[Fig.\,\ref{fig:overviewscheme}\,\textbf{c},\textbf{g}]. We also expect its width to depend on the delay between the two pulses. In contrast to this, the spectral width of the redirected xFID signal is set by the decay time of the emission, which is independent of the delay.

\begin{figure}[tbh]%
\includegraphics[width=\columnwidth]{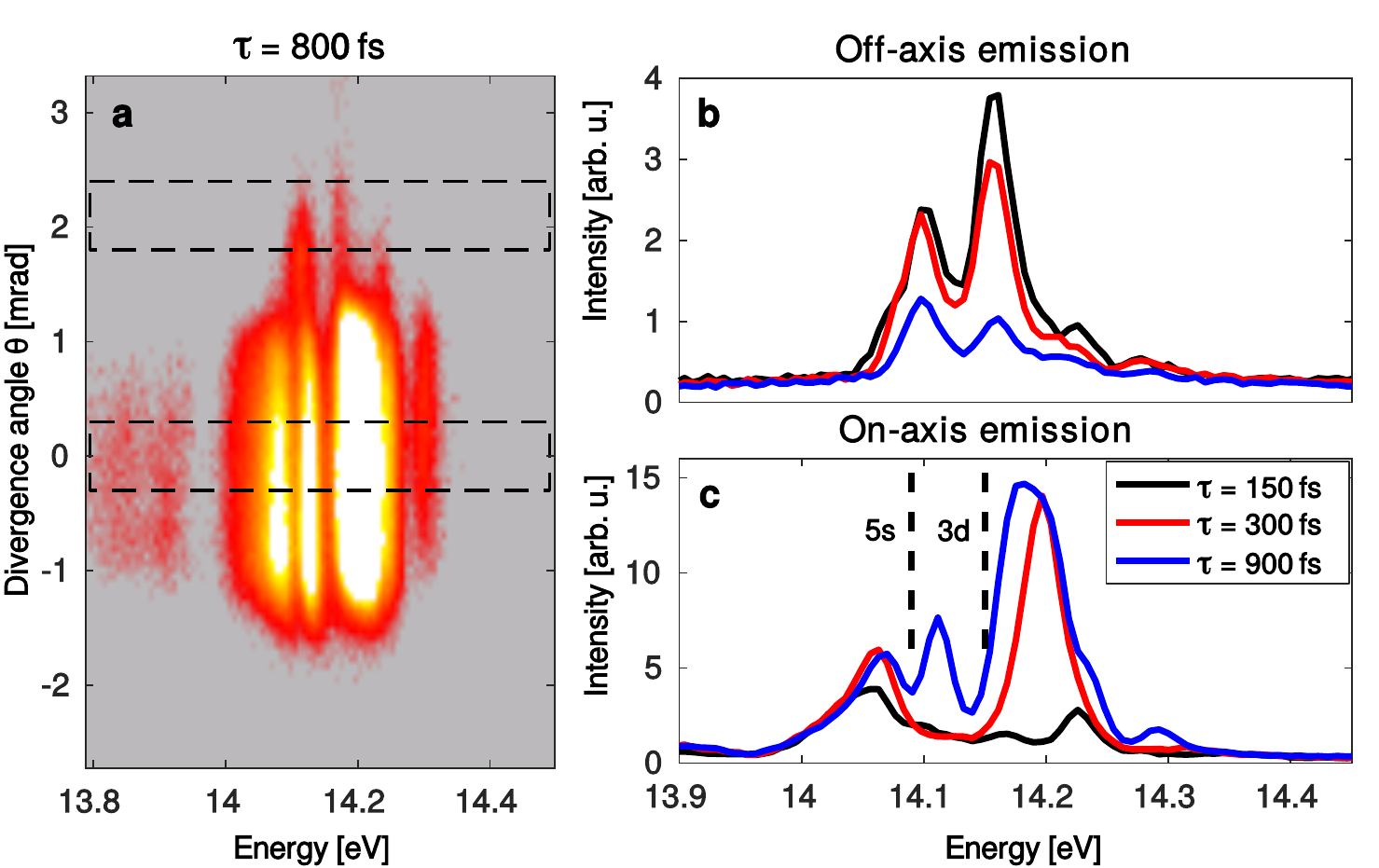}%lbrt2016-06-27 scan4
\caption{\textbf{Delay dependence of the on- and off-axis emission.} \textbf{a} a spatial--spectral profile at a delay of 800\,fs. The dashed lines show where the off-axis (upper) and on-axis (lower) line-out is taken. \textbf{b} A line-out of the off-axis emission that shows the same spectral width for all delays. \textbf{c} A line-out of the on-axis emission for different delays. The dark dashed lines show the energies corresponding to states in argon (5s and 3d). With increasing delay the absorption holes becomes more narrow, and at a delay of 900\,fs one can resolve the two states.}
\label{fig:timepicture}
\end{figure}

In Fig.\,~\ref{fig:timepicture} we investigate these spectral features as a function of the relative delay ($\tau$) between the pump and control pulse. The intensity of the control pulse is reduced as compared with Fig.\,\ref{fig:intensity} in order to minimize the perturbation of the excitation process when the two pulses overlap temporally. 
%To further demonstrate the direction control the xFID emission is here shifted upwards. The redirection is now smaller than in Fig.\,\ref{fig:intensity} but there is still a clear off-axis signal. Comparing the 
In this experiment we direct the xFID emission upwards. Fig.\,~\ref{fig:timepicture} compares the 
spectral width of the on-axis absorption and the xFID off-axis emission and shows that
%, it is clear from Fig.\,~\ref{fig:timepicture} that 
the spectral width of the off-axis emission is not affected by the delay, while the total signal is reduced. In contrast, the spectral width of the on-axis absorption shows an inverse relation with the time delay, in agreement with our description of the process. In the off-axis emission the two states are clearly resolved as emission peaks for all delays, while they can only be resolved as two absorption dips on-axis for delays larger than 500\,fs.

To model these experimental results  we use the theoretical framework outlined in Gaarde~\textit{et al.}~\cite{Gaarde-2011}. We numerically solve the coupled time-dependent Schr\"{o}dinger equation (TDSE), in the single active electron approximation (SAE), and the Maxwell wave equation (MWE). While the experiment considers a non-coaxial geometry for the two pulses, to simplify the computations we use a coaxial geometry in our simulations. The primary difference between these two geometries is that the non-coaxial geometry redirects the resonant XUV emission in a single direction, whereas the coaxial geometry directs the resonant XUV emission into a halo around the shared axis. For the SAE-TDSE calculations we use a pseudopotential that accurately reproduces the singly-excited energy levels of helium. The IR pulse has a central wavelength of $770$\,nm, a FWHM duration of $30.8$\,fs, and a peak intensity of $0.5$\,TW/cm$^{2}$, while the XUV pulse has a central energy of 21.1\,eV, a FWHM duration of $20.5$\,fs, and a peak intensity of $10^{10}$ W/cm$^2$. The central energy of the XUV is chosen to match the $\mathrm{1s^2-1s2p}$ excitation energy in our helium potential. We impose a $360$\,fs dephasing time on the dipole, which, though it is much shorter than the true dephasing time, is much longer than the time scale set by the IR pulse and the relative XUV-IR delays we use. The confocal parameters of the two beams are chosen such that they have similar radial extents, with a $b_{\mathrm{IR}} = 1$\,cm and $b_{\mathrm{XUV}} = 25$\,cm. This allows the atoms that interact with the XUV field at the center of the IR beam to experience a much different Stark shift than the atoms that interact with the XUV field at the edge. To further reduce the required computational time, we restrict the fields to interact with only a single plane of atoms, with a width of $0.001$\,cm and a large atomic density of $5\cdot 10^{18}$\,atoms/cm$^{3}$. The calculated near-field macroscopic electric field is transformed to the far field, and we plot the azimuthally integrated spectral intensity.

\begin{figure}%
\includegraphics[width=\columnwidth,clip]{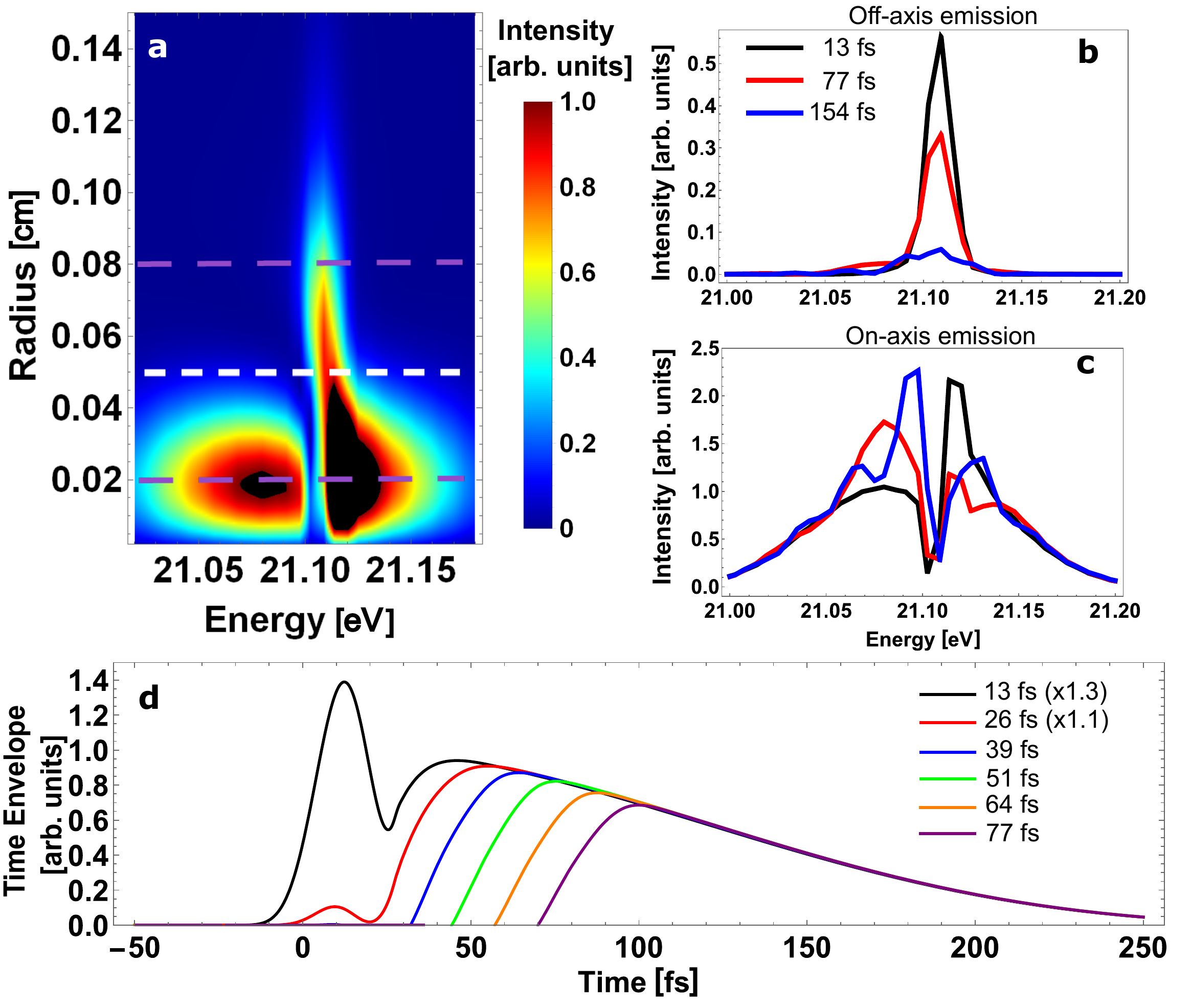}%lbrt
\caption{\textbf{Theroretical calculations.} \textbf{a} Calculated, azimuthally integrated, spectral intensity for an XUV pulse centered on the $\mathrm{1s^2-1s2p}$ excitation energy of helium, with a XUV-IR delay of 13\,fs. Note that black is saturated above the maximum on this color scale. The purple dashed lines show where the off-axis (upper line) and on-axis (lower line) line-out is taken for \textbf{b} and \textbf{c}, respectively. The white dashed line show the lower boundary of what we consider for the off-axis emission signal analyzed in \textbf{d}. \textbf{b} A line-out of the off-axis emission that shows the same spectral width for all delays, compared with a calculation that has no IR pulse. \textbf{c} A line-out of the on-axis emission for different delays, compared with a calculation that has no IR pulse. With increasing delay the profile approaches the no IR case and the absorption lineshape narrows. \textbf{d} shows the radially averaged near field time profile of the off-axis emission for different delay. In these time profile calculations, the XUV pulse is centered at $\mathrm{t=0}$. and the IR pulse is centered at the given delay.}%
\label{fig:theory-fig}%
\end{figure}

The results are shown in Fig.\,\ref{fig:theory-fig}~\textbf{a} for the case of an IR pulse arriving 13\,fs after the XUV pulse. Here we see that near to the common propagation axis, for example at $0.02$\,cm, the XUV shows absorption at the resonant frequency, whereas off-axis, for example at $0.08$\,cm, there is an emission feature at the resonant frequency. In our calculations, this off-axis resonant emission only occurs when the IR pulse overlaps with the XUV pulse in time or arrives after the XUV pulse, which agrees well with the experimental observations. We also find (not shown) that we can reduce the radial extent of this off-axis emission by reducing the peak intensity of the IR pulse, in agreement with the experimental observations illustrated in Fig.\ref{fig:intensity}~\textbf{b}-\textbf{d}. 

To compare our theoretical calculations to these experimental results, we plot lineouts of the azimuthally integrated, spectral intensity at different delays for the off-axis and on-axis case in Fig.\,\ref{fig:theory-fig}~\textbf{b}-\textbf{c}, respectively. 
These off-axis and on-axis results agree well with the observed experimental results in Fig. \ref{fig:timepicture}~\textbf{b}-\textbf{c}.
%Question for Seth: what is meant by the "on-axis case" and the "off-axis case"? 
%Seth response: The on-axis case is the purple line at .02 cm and the off-axis case is the purple line at .08 cm in part a.
For the off-axis case, we find that increasing the IR-XUV delay reduces the intensity of the resonant emission feature, approaching the results for the case of no IR pulse at delays close to the dephasing time of the dipole. The off-axis emission feature maintains a Lorentzian lineshape with a near constant spectral width for all delays. In contrast, the on-axis case shows much less intensity modulation with increasing delay, however, the dispersive lineshape evolves significantly as we vary the IR-XUV delay. The on-axis absorption lineshape becomes narrower with increasing delay, approaching the on-axis results for the case of no IR pulse at long delays.  

\begin{figure}
\includegraphics[width=\columnwidth,clip]{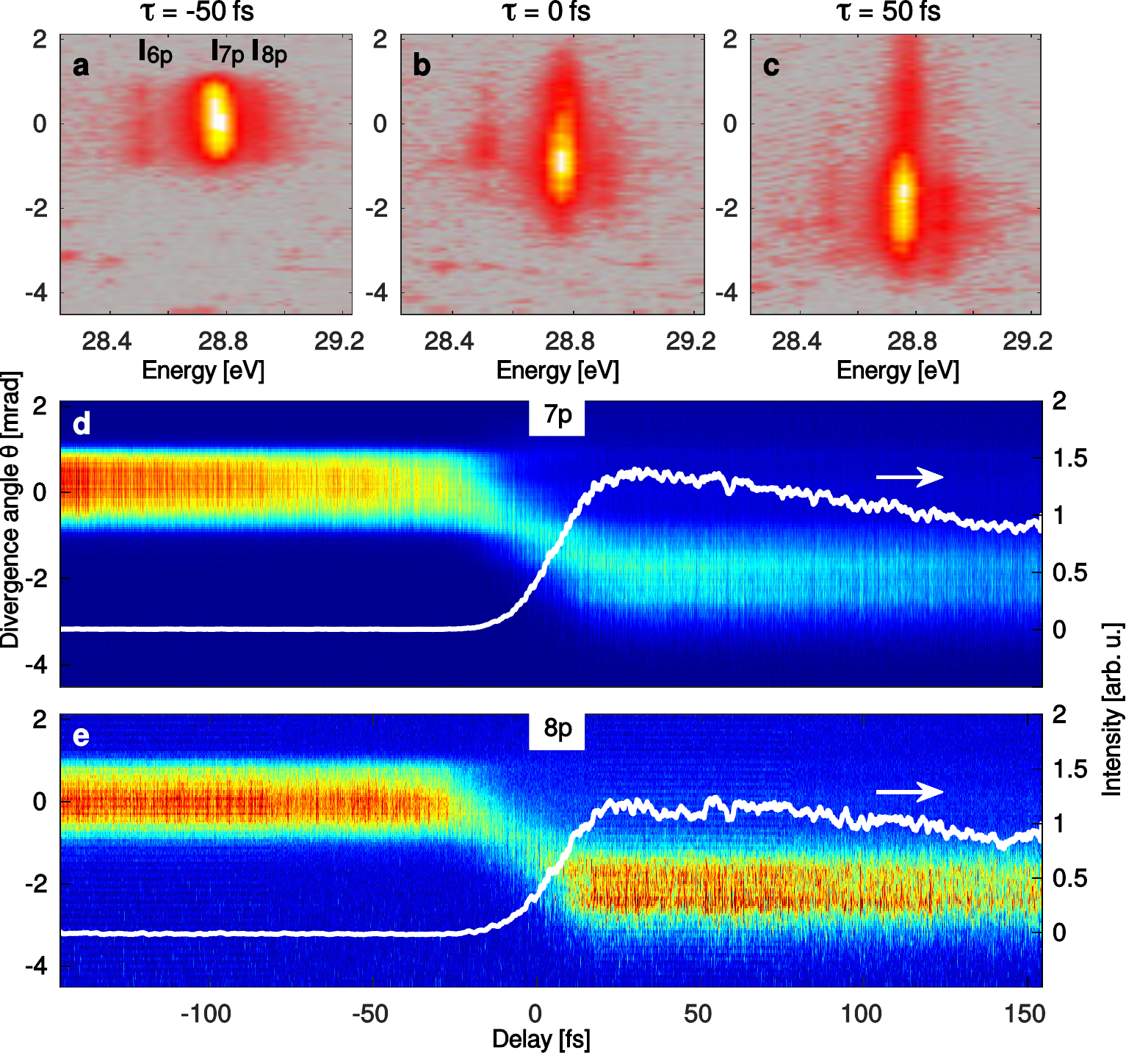}%lbrt 2015-06-25 scan4
\caption{\textbf{Delay dependence of xFID emission} \textbf{a}-\textbf{c} Spatial spectral profile of XUV light transmitted through the argon gas for three different delays between the IR and XUV pulse. Negative delays correspond to the IR pulse preceding the XUV pulse. \textbf{d}-\textbf{e} Lineout of how the emission direction changes as a function of delay from the different excited states (7p and 8p). The signal is normalized to when the IR pulse precedes the XUV pulse. The off-axis signal is integrated around -2\,mrad and shown as a white line, normalized to the end value.}
\label{fig:Spatial}
\end{figure}

Next we experimentally follow the delay dependence of the xFID emission while also showing that our technique can be extended to shorter XUV wavelengths. To this end, we excite several autoionizing states in argon. The 3s$^1$3p$^5$np series of autoionizing states in argon have much higher excitation energies and shorter lifetimes than the states previously presented in this article. We focus on the control of states with principal quantum number n$\geq 6$ since these states can be expected to have  lifetimes longer than the duration of our control pulse~\cite{rubenssonPRL1999,WangPRL2010}. The laser system is tuned to a carrier-wavelength of 820\,nm resulting in harmonic 19 spectrally overlapping with the n = \{6,7,8\} states [The vertical lines in Fig.~\ref{fig:Spatial}~\textbf{a}]. Fig.~\ref{fig:Spatial}\,\textbf{a}-\textbf{c} present the experimentally measured spatial-spectral profile for three different delays between the XUV and IR pulses. When the IR pulse arrives 50\,fs~[Fig.~\ref{fig:Spatial}~\textbf{c}] after the XUV pulse, we observe that the on-axis light is almost fully redirected. The efficiency of the process depends on the conversion efficiency to xFID emission and the efficiency of the control over the xFID signal. Absorbing photons from the XUV pulse and converting them to xFID emission depends on the amount of interacting atoms and thus the gas pressure, which can easily be controlled. The redirection efficiency depends on the shape of the control pulse and can be very high. With the gas pressure and control pulse used in the experiment more than 70\,\% of the emission is redirected.

The temporal dynamics of the wavefront rotation induced by the IR-imposed phase shift is explored further in Fig.~\ref{fig:Spatial}~\textbf{d}-\textbf{e}, where the temporal structure of the redirected xFID emission at the energies corresponding to the n = \{7,8\} states is plotted as function of delay. When the IR pulse precedes the XUV pulse the emission has the same direction and divergence as the pump pulse. In this case the coherence times of the states are short with no time for de-phasing and therefore no increase in the divergence of the xFID emission. When the two pulses start to overlap, the xFID emission begins to change direction. With the intensity gradient used in this experiment the maximum wavefront rotation speed is about 0.06\,mrad/fs. For larger delays the direction no longer changes and the decay can be followed by integrating the off-axis signal, Fig.~\ref{fig:Spatial}~\textbf{d}-\textbf{e}. The decay times for the xFID signals from the 7p and 8p states are found to be $264\pm 2$\,fs and $405\pm 9$\,fs, respectively, assuming a Gaussian temporal profile of the IR pulse.

Our theoretical framework provides us with a way of testing our picture of xFID control by calculating the time profile of the near-field radiation that leads to the observed off-axis signal in the presence of the IR field. As shown in Fig. \ref{fig:theory-fig}\,\textbf{d}, this time profile is calculated by spatially selecting the far field radiation outside a radius of $0.05$\,cm,  then back-transforming to the near field and finally radially integrating the result. For normalization purposes, we then subtract the corresponding time-profile of the off-axis radiation generated in the absence of the IR field. This allows us to directly access the time profile of the off-axis emission feature that results from the radially dependent Stark shift introduced by the IR field. For a 13\,fs IR delay (black), the time profile of the off-axis radiation follows that of the dipole response, showing both a linear response to the XUV until the end of the pulse ($t \approx 30$\,fs) and then a long xFID tail after the end of the XUV pulse. As the XUV-IR delay increases, the time profile of the off-axis emission is seen to result from later in the dipole response, with the time profile always beginning at a time near the peak of the IR field, where the maximum Stark shift occurs. This analysis strongly supports the physical picture presented for the generation of this off-axis resonant emission.

\section*{Summary}
Our experimental and theoretical results all fit well within the  framework presented in the introduction: An  XUV pulse induces a long-lived dipole polarization at a resonant frequency that is out of phase with the driving pulse, and then an IR pulse, arriving at a later time, induces a spatially dependent phase shift on this polarization. The generated signal that occurs before the phase shift continues to propagate forward and appears as absorption when interfering with the driving XUV pulse at the detector. The polarization that occurs during and after this phase shift leads to light propagating off-axis and appearing as new emission, free of the background due to the driving pulse. The redirected emission process is very efficient since almost all the emission from the excited states can be redirected. An additional control pulse would allow for even more precise manipulation of the temporal structure of the emission. 

The advent of pulse shaping in the IR and visible regimes by acousto-optic and electro-optic modulators has proven an invaluable tool in a broad range of a photonic fields, such as quantum control, ultrafast science, and telecommunication. Our work extending this methodology to the XUV by using phase and amplitude control to build an opto-optical modulator opens a similar range of possibilities. For example, using multiple IR control pulses will allow for the rapid switching of xFID emission along a particular direction and thus precise temporal control over narrow-band XUV sources. The ability to control the XUV emission in both time and space will also  make these light source attractive to a wider variety of users, and there are many exciting opportunities ranging from ultrafast quantum information in the XUV to nanoplasmonics studies and femtochemistry. It is also possible to envision the seeding of free-electron lasers with controllable narrow-band XUV light in order to minimize temporal jitter effects~\cite{LambertNP2008}. These same techniques may also help improve the synchronization between light from FELs and lasers by exciting the ensemble of atoms using the FEL and then producing an XUV pulse using opto-optical modulation. The resulting XUV pulse will then be synchronized with the laser pulses used to redirect the emission,  at the  femtosecond level of precision.

\begin{figure*}
\includegraphics[scale=0.5]{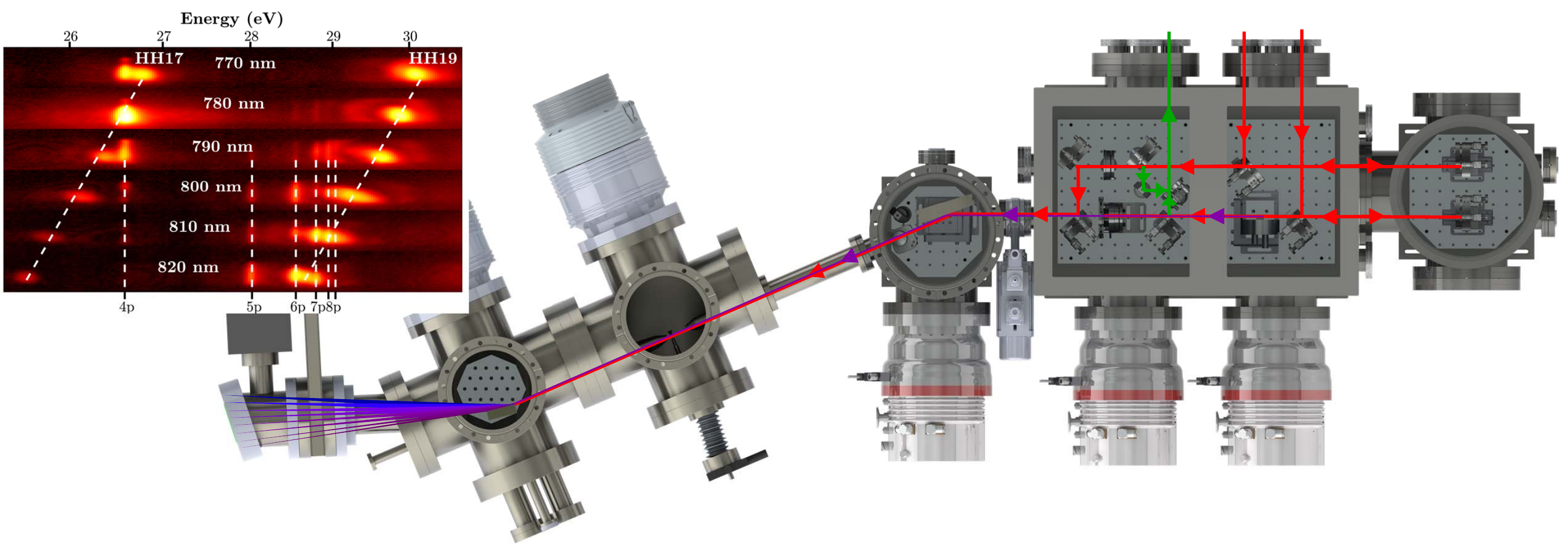}
\caption{\textbf{Overview of the experimental apparatus.} An attosecond pump-probe interferometer is used in a co-propagating non-concentric geometry. The IR beam path is indicated with red lines, the XUV beam path with purple, and the split-off HeNe for active stabilization is indicated with green lines. The top left panel shows the XUV spectrum transmitted through argon gas without the control field, as a function of the carrier wavelength of the driving laser field. Also shown are the Fano energy levels in argon with vertical dashed lines.}
\label{fig:experimentalapparatus}
\end{figure*}

\begin{addendum}
\item
This research was supported by the Swedish Foundation for Strategic Research, the
Marie Curie program ATTOFEL (ITN), the European Research Council
(PALP), the Swedish Research Council, and the Knut and Alice Wallenberg Foundation. Research at LSU supported  by the U.S. Department of Energy, Office of Science, Basic Energy Sciences, under contract no.~DE-FG02-13ER16403.

\item[Competing Interests] The authors declare that they have no
competing financial interests.
\item[Correspondence] Correspondence and requests for materials
should be addressed to J.M.~(~Johan.Mauritsson@fysik.lth.se~).
\item[Authors contributions]
S.B., E.W.L., D.K., L.R. and J.M. designed the experiment and built the experimental apparatus. S.B. and E.W.L. conducted the experiments. A.L.H., C.L.A., M.M. and E.W.L. delivered and maintained the experimental laser system.
%S.B., J.M., E.W.L. and L.R. performed the experimental data analysis and interpretation.
S.C., M.B.G., and K.J.S. carried out the theoretical calculations and contributed to the interpretation of the experimental results.
J.M., E.W.L., S.B. and K.J.S. wrote major parts of the manuscript.
S.B. and E.W.L. contributed equally to this work. All authors contributed to the discussion of the results and commented on the manuscript.
\\
\end{addendum}

\begin{methods}
The experimental setup relies on an amplified titanium sapphire laser system, which delivers 5\,mJ, pulses, with a carrier wavelength around 800\,nm, at a repetition rate of 1\,kHz. The laser system contains two acousto-optical programmable dispersive filters (Dazzler and Mazzler from Fastlite), capable of shaping the spectral phase and amplitude of the amplified pulses. The Mazzler is placed inside the cavity of a regenerative amplifier, where it is used to counteract the effects of gain narrowing by selectively diffracting the spectral components with the highest gain out of the cavity. The use of the Mazzler allows for a top-hat shaped spectral gain-profile of the laser system with a bandwidth of around 100\,nm. The Dazzler is placed in the stretched oscillator beam and is used to pre-compensate any non-linear phase  accumulated by the pulse in the amplifier system in order to generate pulses with near Transform-limited duration ($\mathrm{T}_{\textrm{IR}}\approx 20$\,fs ) after a grating compressor. In addition to optimizing the compression of the pulses the Dazzler is used to restrict the bandwidth of the seed pulses for the laser chain. The combination of the Dazzler and Mazzler allows for a tunability of the carrier wavelength of the laser system to roughly 50\,nm (775\,nm-825\,nm), while maintaining a bandwidth of 50\,nm. The duration of the bandwidth reduced pulses correspond to $\mathrm{T}_{\textrm{IR}} \approx 30$\,fs.

The pulses were directed into a balanced Mach-Zehnder interferometer actively stabilized by a co-propagating frequency stable helium-neon laser [Fig. \ref{fig:experimentalapparatus}]. Both the separation and recombination of the two interferometer arms were done by mirrors with holes drilled through their center. In one of the arms, high-order harmonics were produced by focusing the doughnut-shaped pulses into a pulsed gas cell. A motorized iris placed in the far-field was used to remove the leftover fundamental field of the generation beam. The harmonic and the IR pulses were recombined in a nonconcentric parallel propagating geometry and refocused into a second pulsed gas jet using a toroidal mirror. Light transmitted through the second cell was then sent into a home-built flat-field imaging spectrometer based on a variable-line-spacing grating and a microchannel-plate with
an attached phosphor screen and a camera with a resolution of 2456x2058 pixels and a dynamic range of 14 bits~\cite{LorekRSI2014}.
\end{methods}

\begin{addendum}
\item[Data availability.]
The data that support the findings of this study are available from the corresponding author upon reasonable request. 
\end{addendum}

\bibliographystyle{naturemag}

\bibliography{bibliography}

\begin{thebibliography}{10}
\expandafter\ifx\csname url\endcsname\relax
  \def\url#1{\texttt{#1}}\fi
\expandafter\ifx\csname urlprefix\endcsname\relax\def\urlprefix{URL }\fi
\providecommand{\bibinfo}[2]{#2}
\providecommand{\eprint}[2][]{\url{#2}}

\bibitem{DebyePNAS1932}
\bibinfo{author}{{Debye}, P.} \& \bibinfo{author}{{Sears}, F.~W.}
\newblock \bibinfo{title}{{On the Scattering of Light by Supersonic Waves}}.
\newblock \emph{\bibinfo{journal}{Proceedings of the National Academy of
  Science}} \textbf{\bibinfo{volume}{18}}, \bibinfo{pages}{409--414}
  (\bibinfo{year}{1932}).

\bibitem{weinerRSI2000}
\bibinfo{author}{Weiner, A.~M.}
\newblock \bibinfo{title}{Femtosecond pulse shaping using spatial light
  modulators}.
\newblock \emph{\bibinfo{journal}{Review of Scientific Instruments}}
  \textbf{\bibinfo{volume}{71}}, \bibinfo{pages}{1929--1960}
  (\bibinfo{year}{2000}).

\bibitem{KaoRMP2010}
\bibinfo{author}{Kao, C.~K.}
\newblock \bibinfo{title}{Nobel lecture: Sand from centuries past: Send future
  voices fast}.
\newblock \emph{\bibinfo{journal}{Rev. Mod. Phys.}}
  \textbf{\bibinfo{volume}{82}}, \bibinfo{pages}{2299} (\bibinfo{year}{2010}).

\bibitem{PaulScience2001}
\bibinfo{author}{Paul, P.~M.} \emph{et~al.}
\newblock \bibinfo{title}{Observation of a train of attosecond pulses from high
  harmonic generation}.
\newblock \emph{\bibinfo{journal}{Science}} \textbf{\bibinfo{volume}{292}},
  \bibinfo{pages}{1689} (\bibinfo{year}{2001}).

\bibitem{HentschelNature2001}
\bibinfo{author}{Hentschel, M.} \emph{et~al.}
\newblock \bibinfo{title}{Attosecond metrology}.
\newblock \emph{\bibinfo{journal}{Nature}}
  \textbf{\bibinfo{volume}{\textbf{414}}}, \bibinfo{pages}{509}
  (\bibinfo{year}{2001}).

\bibitem{KrauszRMP2009}
\bibinfo{author}{Krausz, F.} \& \bibinfo{author}{Ivanov, M.}
\newblock \bibinfo{title}{Attosecond physics}.
\newblock \emph{\bibinfo{journal}{Rev. Mod. Phys.}}
  \textbf{\bibinfo{volume}{81}}, \bibinfo{pages}{163--234}
  (\bibinfo{year}{2009}).

\bibitem{McPhersonJOSAB1987}
\bibinfo{author}{McPherson, A.} \emph{et~al.}
\newblock \bibinfo{title}{Studies of multiphoton production of
  vacuum-ultraviolet radiation in the rare gases}.
\newblock \emph{\bibinfo{journal}{J. {O}pt. {S}oc. {A}m.~{B}}}
  \textbf{\bibinfo{volume}{\textbf{4}}}, \bibinfo{pages}{595}
  (\bibinfo{year}{1987}).

\bibitem{FerrayJPB1988}
\bibinfo{author}{Ferray, M.}, \bibinfo{author}{L'Huillier, A.},
  \bibinfo{author}{Li, X.~F.}, \bibinfo{author}{Mainfray, G.} \&
  \bibinfo{author}{Manus, C.}
\newblock \bibinfo{title}{Multiple-harmonic conversion of 1064 nm radiation in
  rare gases}.
\newblock \emph{\bibinfo{journal}{J. Phys. B: At., Mol. Opt. Phys.}}
  \textbf{\bibinfo{volume}{\textbf{21}}}, \bibinfo{pages}{L31}
  (\bibinfo{year}{1988}).

\bibitem{SchaferPRL1993}
\bibinfo{author}{Schafer, K.~J.}, \bibinfo{author}{Yang, B.},
  \bibinfo{author}{DiMauro, L.~F.} \& \bibinfo{author}{Kulander, K.~C.}
\newblock \bibinfo{title}{Above threshold ionization beyond the high harmonic
  cutoff}.
\newblock \emph{\bibinfo{journal}{Phys. {R}ev. {L}ett.}}
  \textbf{\bibinfo{volume}{\textbf{70}}}, \bibinfo{pages}{1599}
  (\bibinfo{year}{1993}).

\bibitem{CorkumPRL1993}
\bibinfo{author}{Corkum, P.~B.}
\newblock \bibinfo{title}{Plasma perspective on strong-field multiphoton
  ionization}.
\newblock \emph{\bibinfo{journal}{Phys. {R}ev. {L}ett.}}
  \textbf{\bibinfo{volume}{\textbf{71}}}, \bibinfo{pages}{1994}
  (\bibinfo{year}{1993}).

\bibitem{LewensteinPRA1994}
\bibinfo{author}{Lewenstein, M.}, \bibinfo{author}{Balcou, P.},
  \bibinfo{author}{Ivanov, M.}, \bibinfo{author}{L'Huillier, A.} \&
  \bibinfo{author}{Corkum, P.~B.}
\newblock \bibinfo{title}{Theory of high-order harmonic generation by
  low-frequency laser fields}.
\newblock \emph{\bibinfo{journal}{Phys. {R}ev. {A}}}
  \textbf{\bibinfo{volume}{49}}, \bibinfo{pages}{2117} (\bibinfo{year}{1994}).

\bibitem{AckermannNP2007}
\bibinfo{author}{W.~Ackermann, V. A. e.~a., G.~Asova}.
\newblock \bibinfo{title}{Operation of a free-electron laser from the extreme
  ultraviolet to the water window}.
\newblock \emph{\bibinfo{journal}{Nat. Photon.}} \textbf{\bibinfo{volume}{1}},
  \bibinfo{pages}{336} (\bibinfo{year}{2007}).

\bibitem{EmmaNP2010}
\bibinfo{author}{Emma, P.} \emph{et~al.}
\newblock \bibinfo{title}{First lasing and operation of an angstrom-wavelength
  free-electron laser}.
\newblock \emph{\bibinfo{journal}{Nat. Photon.}} \textbf{\bibinfo{volume}{4}},
  \bibinfo{pages}{641} (\bibinfo{year}{2010}).

\bibitem{DrescherNature2002}
\bibinfo{author}{Drescher, M.} \emph{et~al.}
\newblock \bibinfo{title}{Time-resolved atomic inner-shell spectroscopy}.
\newblock \emph{\bibinfo{journal}{Nature}} \textbf{\bibinfo{volume}{419}},
  \bibinfo{pages}{803--807} (\bibinfo{year}{2002}).

\bibitem{KlingScience2006}
\bibinfo{author}{Kling, M.~F.} \emph{et~al.}
\newblock \bibinfo{title}{Control of electron localization in molecular
  dissociation}.
\newblock \emph{\bibinfo{journal}{Science}} \textbf{\bibinfo{volume}{312}},
  \bibinfo{pages}{246} (\bibinfo{year}{2006}).

\bibitem{StockmanNPho2007}
\bibinfo{author}{Stockman, M.}, \bibinfo{author}{Kling, M.},
  \bibinfo{author}{Kleineberg, U.} \& \bibinfo{author}{Krausz, F.}
\newblock \bibinfo{title}{Attosecond nanoplasmonic-field microscope}.
\newblock \emph{\bibinfo{journal}{Nature Photonics}}
  \textbf{\bibinfo{volume}{1}}, \bibinfo{pages}{539--544}
  (\bibinfo{year}{2007}).

\bibitem{CavalieriNature2007}
\bibinfo{author}{Cavalieri, A.~L.} \emph{et~al.}
\newblock \bibinfo{title}{Attosecond spectroscopy in condensed matter}.
\newblock \emph{\bibinfo{journal}{Nature}} \textbf{\bibinfo{volume}{449}},
  \bibinfo{pages}{1029} (\bibinfo{year}{2007}).

\bibitem{SansoneNat2010}
\bibinfo{author}{Sansone, G.} \emph{et~al.}
\newblock \bibinfo{title}{Electron localization following attosecond molecular
  photoionization}.
\newblock \emph{\bibinfo{journal}{Nature}} \textbf{\bibinfo{volume}{465}},
  \bibinfo{pages}{763} (\bibinfo{year}{2010}).

\bibitem{FangPRL2010}
\bibinfo{author}{Fang, L.} \emph{et~al.}
\newblock \bibinfo{title}{Double core-hole production in n2: Beating the auger
  clock}.
\newblock \emph{\bibinfo{journal}{Phys. Rev. Lett.}}
  \textbf{\bibinfo{volume}{105}}, \bibinfo{pages}{083005}
  (\bibinfo{year}{2010}).

\bibitem{CryanPRL2010}
\bibinfo{author}{Cryan, J.~P.} \emph{et~al.}
\newblock \bibinfo{title}{Auger electron angular distribution of double
  core-hole states in the molecular reference frame}.
\newblock \emph{\bibinfo{journal}{Phys. Rev. Lett.}}
  \textbf{\bibinfo{volume}{105}}, \bibinfo{pages}{083044}
  (\bibinfo{year}{2010}).

\bibitem{GlowniaOE2010}
\bibinfo{author}{Glownia, J.~M.} \emph{et~al.}
\newblock \bibinfo{title}{Time-resolved pump-probe experiments at the lcls}.
\newblock \emph{\bibinfo{journal}{Opt. Exp.}} \bibinfo{pages}{017620}
  (\bibinfo{year}{2010}).

\bibitem{YoungNat2010}
\bibinfo{author}{Young, L.} \emph{et~al.}
\newblock \bibinfo{title}{Femtosecond electronic response of atoms to
  ultra-intense x-rays}.
\newblock \emph{\bibinfo{journal}{Nature}} \textbf{\bibinfo{volume}{466}},
  \bibinfo{pages}{56} (\bibinfo{year}{2010}).

\bibitem{GoulielmakisNat2010}
\bibinfo{author}{Goulielmakis, E.} \emph{et~al.}
\newblock \bibinfo{title}{Real-time observation of valence electron motion}.
\newblock \emph{\bibinfo{journal}{Nature}} \textbf{\bibinfo{volume}{466}},
  \bibinfo{pages}{739--743} (\bibinfo{year}{2010}).

\bibitem{DoumyPRL2011}
\bibinfo{author}{Doumy, G.} \emph{et~al.}
\newblock \bibinfo{title}{Nonlinear atomic response to intense ultrashort x
  rays}.
\newblock \emph{\bibinfo{journal}{Phys. {R}ev. {L}ett.}}
  \textbf{\bibinfo{volume}{106}}, \bibinfo{pages}{083002}
  (\bibinfo{year}{2011}).

\bibitem{klunderPRL2011}
\bibinfo{author}{Kl\"under, K.} \emph{et~al.}
\newblock \bibinfo{title}{Probing single-photon ionization on the attosecond
  time scale}.
\newblock \emph{\bibinfo{journal}{Phys. Rev. Lett.}}
  \textbf{\bibinfo{volume}{106}}, \bibinfo{pages}{143002}
  (\bibinfo{year}{2011}).

\bibitem{BlochPR1946}
\bibinfo{author}{Bloch, F.}
\newblock \bibinfo{title}{Nuclear induction}.
\newblock \emph{\bibinfo{journal}{Phys. Rev.}} \textbf{\bibinfo{volume}{70}},
  \bibinfo{pages}{460--474} (\bibinfo{year}{1946}).

\bibitem{HahnPR1950}
\bibinfo{author}{Hahn, E.~L.}
\newblock \bibinfo{title}{Nuclear induction due to free larmor precession}.
\newblock \emph{\bibinfo{journal}{Phys. Rev.}} \textbf{\bibinfo{volume}{77}},
  \bibinfo{pages}{297--298} (\bibinfo{year}{1950}).

\bibitem{brewerPRA1972}
\bibinfo{author}{Brewer, R.~G.} \& \bibinfo{author}{Shoemaker, R.~L.}
\newblock \bibinfo{title}{Optical free induction decay}.
\newblock \emph{\bibinfo{journal}{Phys. Rev. A}} \textbf{\bibinfo{volume}{6}},
  \bibinfo{pages}{2001--2007} (\bibinfo{year}{1972}).

\bibitem{HopfPRA1973}
\bibinfo{author}{Hopf, F.~A.}, \bibinfo{author}{Shea, R.~F.} \&
  \bibinfo{author}{Scully, M.~O.}
\newblock \bibinfo{title}{Theory of optical free-induction decay and two-photon
  superradiance}.
\newblock \emph{\bibinfo{journal}{Phys. Rev. A}} \textbf{\bibinfo{volume}{7}},
  \bibinfo{pages}{2105--2110} (\bibinfo{year}{1973}).

\bibitem{Fano1961PR}
\bibinfo{author}{Fano, U.}
\newblock \bibinfo{title}{Effects of configuration interaction on intensities
  and phase shifts}.
\newblock \emph{\bibinfo{journal}{Phys. {R}ev.}}
  \textbf{\bibinfo{volume}{124}}, \bibinfo{pages}{1866--1878}
  (\bibinfo{year}{1961}).

\bibitem{WangPRL2010}
\bibinfo{author}{Wang, H.} \emph{et~al.}
\newblock \bibinfo{title}{Attosecond time-resolved autoionization of argon}.
\newblock \emph{\bibinfo{journal}{Phys. {R}ev. {L}ett.}}
  \textbf{\bibinfo{volume}{105}}, \bibinfo{pages}{143002}
  (\bibinfo{year}{2010}).

\bibitem{OttNature2014}
\bibinfo{author}{Ott, C.} \emph{et~al.}
\newblock \bibinfo{title}{Reconstruction and control of a time-dependent
  two-electron wave packet}.
\newblock \emph{\bibinfo{journal}{Nature}} \textbf{\bibinfo{volume}{516}},
  \bibinfo{pages}{374--378} (\bibinfo{year}{2014}).

\bibitem{BeckNJP2014}
\bibinfo{author}{Beck, A.~R.} \emph{et~al.}
\newblock \bibinfo{title}{Attosecond transient absorption probing of electronic
  superpositions of bound states in neon: detection of quantum beats}.
\newblock \emph{\bibinfo{journal}{New Journal of Physics}}
  \textbf{\bibinfo{volume}{16}}, \bibinfo{pages}{113016}
  (\bibinfo{year}{2014}).

\bibitem{LiaoPRL2015}
\bibinfo{author}{Liao, C.-T.}, \bibinfo{author}{Sandhu, A.},
  \bibinfo{author}{Camp, S.}, \bibinfo{author}{Schafer, K.~J.} \&
  \bibinfo{author}{Gaarde, M.~B.}
\newblock \bibinfo{title}{Beyond the single-atom response in absorption line
  shapes: Probing a dense, laser-dressed helium gas with attosecond pulse
  trains}.
\newblock \emph{\bibinfo{journal}{Phys. Rev. Lett.}}
  \textbf{\bibinfo{volume}{114}}, \bibinfo{pages}{143002}
  (\bibinfo{year}{2015}).

\bibitem{wuPRA2013}
\bibinfo{author}{Wu, M.}, \bibinfo{author}{Chen, S.}, \bibinfo{author}{Gaarde,
  M.~B.} \& \bibinfo{author}{Schafer, K.~J.}
\newblock \bibinfo{title}{Time-domain perspective on autler-townes splitting in
  attosecond transient absorption of laser-dressed helium atoms}.
\newblock \emph{\bibinfo{journal}{Phys. Rev. A}} \textbf{\bibinfo{volume}{88}},
  \bibinfo{pages}{043416} (\bibinfo{year}{2013}).

\bibitem{Vura-weisPCL2013}
\bibinfo{author}{Vura-Weis, J.} \emph{et~al.}
\newblock \bibinfo{title}{Femtosecond m$_{2,3}$-edge spectroscopy of
  transition-metal oxides: Photoinduced oxidation state change in
  $\alpha$-fe$_2$o$_3$}.
\newblock \emph{\bibinfo{journal}{The Journal of Physical Chemistry Letters}}
  \textbf{\bibinfo{volume}{4}}, \bibinfo{pages}{3667--3671}
  (\bibinfo{year}{2013}).

\bibitem{BeaulieuPRL2016}
\bibinfo{author}{Beaulieu, S.} \emph{et~al.}
\newblock \bibinfo{title}{Role of excited states in high-order harmonic
  generation}.
\newblock \emph{\bibinfo{journal}{Phys. Rev. Lett.}}
  \textbf{\bibinfo{volume}{117}}, \bibinfo{pages}{203001}
  (\bibinfo{year}{2016}).

\bibitem{OttScience2013}
\bibinfo{author}{Ott, C.} \emph{et~al.}
\newblock \bibinfo{title}{Lorentz meets fano in spectral line shapes: A
  universal phase and its laser control}.
\newblock \emph{\bibinfo{journal}{Science}} \textbf{\bibinfo{volume}{340}},
  \bibinfo{pages}{716--720} (\bibinfo{year}{2013}).

\bibitem{Gaarde-2011}
\bibinfo{author}{Gaarde, M.~B.}, \bibinfo{author}{Buth, C.},
  \bibinfo{author}{Tate, J.~L.} \& \bibinfo{author}{Schafer, K.~J.}
\newblock \bibinfo{title}{Transient absorption and reshaping of ultrafast xuv
  light by laser-dressed helium}.
\newblock \emph{\bibinfo{journal}{Phys. Rev. A}} \textbf{\bibinfo{volume}{83}},
  \bibinfo{pages}{013419} (\bibinfo{year}{2011}).

\bibitem{rubenssonPRL1999}
\bibinfo{author}{Rubensson, J.-E.} \emph{et~al.}
\newblock \bibinfo{title}{Influence of the radiative decay on the cross section
  for double excitations in helium}.
\newblock \emph{\bibinfo{journal}{Phys. Rev. Lett.}}
  \textbf{\bibinfo{volume}{83}}, \bibinfo{pages}{947--950}
  (\bibinfo{year}{1999}).

\bibitem{LambertNP2008}
\bibinfo{author}{Lambert, G.} \emph{et~al.}
\newblock \bibinfo{title}{Injection of harmonics generated in gas in a
  free-electron laser providing intense and coherent extreme-ultraviolet
  light}.
\newblock \emph{\bibinfo{journal}{Nature Physics}}
  \textbf{\bibinfo{volume}{4}}, \bibinfo{pages}{296--300}
  (\bibinfo{year}{2008}).

\bibitem{LorekRSI2014}
\bibinfo{author}{Lorek, E.} \emph{et~al.}
\newblock \bibinfo{title}{High-order harmonic generation using a
  high-repetition-rate turnkey laser}.
\newblock \emph{\bibinfo{journal}{Rev. Sci. Instrum.}}
  \textbf{\bibinfo{volume}{85}}, \bibinfo{pages}{123106--1--5}
  (\bibinfo{year}{2014}).

\end{thebibliography}

\end{document}